\documentclass[12pt]{article}
\usepackage{authblk}
\usepackage{amsfonts}
\usepackage{amsmath}
\usepackage{amssymb}
\usepackage{amsthm}
\usepackage[utf8]{inputenc}
\usepackage[russian, english]{babel}
\usepackage{color, hyperref}
\newtheorem*{theorem}{Theorem}
\newtheorem{lemma}{Lemma}

\newtheorem{definition}{Definition}

\newtheorem{remark}{Remark}
\newtheorem{corollary}{Corollary}

\newcommand{\Tr}{\mathop{\mathrm{Tr}}\nolimits}

\title{On the Wehrl entropy lower bound for a locally compact abelian group}
 
\author{Evgeny I. Zelenov}
\affil{Steklov Mathematical Institute\\ Russian Academy of Sciences\\
8 Gubkina St., Moscow 119991, Russia}
\date{June, 2023}
\begin{document}
\maketitle

\begin{abstract}
A Wehrl entropy construction is proposed for an arbitrary locally compact abelian group $G$. It is proved that the Wehrl entropy is not less than a non-negative integer, which is an invariant of the group $G$. The minimum of the Wehrl entropy is achieved on coherent states.
\end{abstract}

\section{Introduction: Wehrl's entropy.}
In \cite{Wehrl1, Wehrl2} A.\, Wehrl proposed the following definition of classical entropy corresponding to a quantum system. 

Let $z=(q,p)\in\mathbb R^2$. In the space $L^2(\mathbb R)$ we define a vector (coherent state) by the formula
$$
|z\rangle = \pi^{-1/4}\exp\left(-(x-q)^2/2+ipx\right), \,x\in\mathbb R.
$$
For the density matrix $\rho$, we define the Husimi function $Q_\rho(z) = \langle z|\rho|z\rangle$ and the Wehrl entropy $$S^W(\rho) = -\int\frac{dz}{\pi}Q_\rho(z)\log Q_\rho(z).$$

Wehrl showed that $S^W(\rho)\geq S(\rho) = -\Tr\rho\log\rho$, and the monotonicity property is satisfied. That is, if $\rho_{12}$ is the density matrix of a composite system in the space $L^2(\mathbb R)\otimes L^2(\mathbb R)$, then $S^W(\rho_{12})\geq S^W(\rho_1)$, where $\rho_1 = \Tr_2\rho_{12}$.

In fact, the Wehrl entropy has a stronger property. Let $\rho_{12}$ be the density operator in the space $\mathcal H_1\otimes \mathcal H_2$ and $\rho_1,\,\rho_2$ be partial traces over the spaces $\mathcal H_2$ and $\mathcal H_1$, respectively, then the inequality holds
$$
S^W(\rho_{12})\geq S^W(\rho_1)+S^W(\rho_2) + S(\rho_{12}) - S(\rho_1) - S(\rho_2), 
$$
and equality is achieved when $\rho_{12}=\rho_1\otimes \rho_2$ \cite{LiebSeiringer}.

Wehrl also formulated a conjecture about the minimum of $S^W$, namely, $S^W\geq 1$, and the minimum is achieved on coherent states $|z\rangle\langle z|$, the conjecture was proved in \cite{Lieb}. It follows from the subadditivity of the Wehrl entropy that for the case of $L^2(\mathbb R^N)$ the minimum of entropy is equal to $N$, \cite{Lieb}.

The proof used results on "Gaussian maximizers" for Gaussian kernels \cite{LiebGauss}. The problem of Gaussian maximizers has received a deep generalization in quantum information theory \cite{Holevo}.

The purpose of this paper is to define the Wehrl entropy $S^W_G$ for an arbitrary locally compact abelian group $G$ and to investigate its properties. The main result of the article consists in proving the inequality $S^W_G\geq n=n(G)$, where $n(G)$ is an invariant of the group $G$, while equality is achieved on pure coherent states. 

Everywhere else, the group $G$ is assumed to be Hausdorff.

\section{Representations of CCR.}
The representation theory of canonical commutation relations (CCR) over locally compact abelian groups has a long history and numerous applications, for example, \cite{Rosenberg, Weil, Mackey}. Recently, interest in CCR representations over locally compact groups has arisen in connection with applications to quantum information theory, \cite{Amosov1, Amosov2}.

Let $G$ be a locally compact abelian group. We will use an additive notation for $G$, we will denote the neutral element by $0$.

By $\hat G$ we denote the Pontryagin dual group. We will write $\hat G$ multiplicatively, we will denote the neutral element by $1$. Let $F=G\times\hat G$ and 
$\mathbb T$ is a unit circle in the complex plane.

On $F$ we define the cocycle $\omega\colon F\times F\to\mathbb T$ by the formula:
$$
\omega\big((g,\lambda),(g’,\lambda’)\big) = \lambda(g’)\overline{\lambda’(g)}, \,g, g'\in G,\,\lambda, \lambda'\in\hat G.
$$ 

\begin{definition}
The Heisenberg group $Heis(G)$ is called the central extension of the group $F$ corresponding to the cocycle $\omega$:
$$
Heis(G)= F\times\mathbb T, 
$$
a group operation is defined by the expression
$$
(g,\lambda, t)(g’,\lambda’,t’)=\big(g+g,\lambda\lambda’, tt’\omega\left((g,\lambda),(g’,\lambda’)\right)\big). 
$$	
\end{definition}

\begin{definition}
The Weyl system $(G,\mathcal H)$ is a strongly continuous mapping from $F$ to a set of unitary operators on a complex Hilbert space $\mathcal H$ satisfying the Weyl commutation relations:
$$
W(g,\lambda)W(g’,\lambda’)=\omega\left((g,\lambda),(g’,\lambda’)\right) W(g’,\lambda’)W(g,\lambda).
$$
\end{definition}
This is the same as the projective representation of $F$ defined by the cocycle $\omega$, or the representation of the group $Heis(G)$ identical at the center of this group, or the representation of canonical commutation relations (CCR) in Weyl form. 

The irreducible representation of CCR is unique up to the unitary equivalence \cite{Mackey, Prasad}. 
As a rule, the case of separable Hilbert space $\mathcal H$ is interesting for applications. To do this, it is necessary that the group $G$ has a countable base, we will assume this everywhere else, although this restriction is of a technical nature.

There is a natural representation of CCR in the Hilbert space $H=L^2(G)$
$$
\left(W(g,\lambda )f\right)(h)=\lambda (h)f(h-g),\ g,h\in G,\ \lambda \in \hat G.
$$

Let $(W, \mathcal H)$ be an irreducible representation of CCR. Let's choose an arbitrary vector $|\phi\rangle\in\mathcal H$ of the unit norm. Like any nonzero vector from $\mathcal H$, this vector is cyclic due to the irreducibility of the representation.

\begin{definition}
Set of vectors 
$$
|g,\lambda\rangle = W(g,\lambda)|\phi\rangle,\,\, (g,\lambda)\in F
$$
is called a set of (generalized) coherent states.
\end{definition}

\begin{remark}
The construction depends on the choice of the vector $|\phi\rangle$. If the vector is selected in some special way (vacuum vector), then the corresponding system is called a system of coherent states (not generalized).
\end{remark}

\section{The lower bound of the Wehrl entropy.}

For $z=(g,\lambda)\in F$ by $|z\rangle$ we denote the corresponding coherent state. The Haar measure on $F$ is denoted by $dz$.

The set of one-dimensional projectors $\{|z\rangle \langle z|, \,\,z\in F\}$ forms a decomposition of unity in $\mathcal H$. This is a consequence of the irreducibility of the CCR representation.

The measuring channel is associated with this decomposition of the unit
$$
\rho\to \Phi[\rho]=\int_F\langle z|\rho |z\rangle |z\rangle\langle z|dz.
$$
Denote by $Q_\rho(z) = \langle z|\rho |z\rangle$ the Husimi function of the state $\rho$. The Wehrl entropy is given by the expression:
$$
S^W_G(\rho) = -\int_F Q_\rho(z)\log Q_\rho(z)dz.
 $$

An important characteristic of a coherent state system is the lower bound for the Wehrl entropy. The following results are known:
\begin{itemize}
\item
$G=\mathbb R^N$, then $N\leq S^W_G(\rho)$, and equality is achieved if and only if $\rho$ is a coherent state, \cite{Lieb}.
\item
$G=\mathbb Q_p$, then $0\leq S^W_G(\rho)$, and equality is achieved if and only if $\rho$ is a coherent state, \cite{Zelenov}, here $\mathbb Q_p$ is an additive group of the field of $p$-adic numbers.
\end{itemize}

The main result of this paper is the following theorem.
\begin{theorem}
Let $G$ be a locally compact abelian group. Then the inequality is true
$$
S^W_G\geq n,
$$
where $n=n(G)$ is a non-negative integer that is an invariant of the group $G$.
\end{theorem}

First of all, let's use the Pontryagin-van Kampen structural theorem (\cite{HewittRoss}, theorem (24.30)): any locally compact abelian group is topologically isomorphic to the group $\mathbb R^n\times G_0$, where $G_0$ is some locally compact abelian group containing an open compact subgroup. Moreover, $n=n(G)$ is an invariant of the group $G$ in the following sense. If $G=\mathbb R^m\times G_1$, where $G_1$ is a locally compact abelian group also containing an open compact subgroup, then $m=n$.

We prove  Theorem for the group $G$ containing an open compact subgroup. 
\begin{remark} 
Note that an open subgroup in a topological group is always closed. However, an arbitrary closed subgroup is usually not open. For example, if $G$ is a locally compact abelian group, each nontrivial subgroup of which is open, then it is either a group $\mathbb Z_p$ of $p$-adic integers (in the compact case), or an additive group of the field $\mathbb Q_p$ of $p$-adic numbers (\cite{Robertson}). In other words, the existence of a nontrivial open compact subgroup is a very restrictive condition.
\end{remark}
Let $H$ be an open compact subgroup in $G$.
By $A(\hat G, H)$ we denote the annihilator of the group $H$ in $\hat G$, that is, a subset in $\hat G$ consisting of all such $\lambda$ that $\lambda(H)=1$, $A(\hat G, H)$ -- a subgroup in $\hat G$. Since, by the condition of Theorem, the subgroup $H$ is open, then the quotient group $G/H$ is discrete. At the same time, $A(\hat G, H) = \widehat{G/H}$ is compact. Thus, the group $K=H\times A(\hat G, H)$ is a compact subgroup in $F=G\times\hat G$. Since $H$ is compact, then $A(\hat G, H)$ is an open subgroup in $\hat G$(\cite{HewittRoss}, (23.24d)) and, consequently, the group $F/K$ is discrete.

Note that the cocycle $\omega$ is identically equal to 1 on the group $K$.

This subgroup has the maximality property in the following sense. For every $g\in G, \,g\notin H$ there is such a character $\lambda\in A(\hat G, H)$ that $\lambda(g)\neq 1$. For this reason, the group $K$ will be called a maximal abelian compact subgroup in the group $F$.

Let's prove the following lemma.

\begin{lemma}
Let $K$ be the maximal compact abelian subgroup in $F$. By $|0\rangle$ we denote a vector (of unit norm)  invariant with respect to the action of the operators $W(u), \,u\in K$,
$$
W(u)|0\rangle = |0\rangle\,\,\forall u\in K.
$$
Then the equalities are valid:
\begin{equation}
\label{1}
|\langle g,\lambda |g’,\lambda’\rangle | = 
\begin{cases}
1,&\text{if $(g-g’,\lambda\overline{\lambda’})\in K$},\\
0,&\text{otherwise}.
\end{cases}
\end{equation}	
\end{lemma}

Lemma 1 is of interest because it describes the structure of the set of coherent states for groups with an open compact subgroup. Essentially, the statement of Lemma 1 is as follows: two coherent states $|z\rangle, \,|z'\rangle$ are either orthogonal (if $z$ and $z'$ lie in different cosets in $F/K$) or differ only by a phase factor (otherwise case). Thus, the set of vectors $\{|\alpha\rangle, \,\alpha\in F/K\}$ ($\alpha$ runs through the set of representatives of cosets in $F/K$, one representative from each coset) forms an orthonormal basis in the representation space $\mathcal H$. 

The operators $W(z),\,z\in K$ form a unitary representation of the compact group $K$ in $\mathcal H$. Such a representation always has an invariant vector (of the unit norm), which we denoted by $|0\rangle$. As we will see later, such a vector is unique up to the phase factor.

It is convenient to use the notation $+$ for a group operation in $G\times\hat G$ and $-$  for the inverse operation, that is, $z+z'=(g+g',\lambda\lambda'), \,z-z'=(g-g',\lambda\overline{\lambda'})$.  Taking into account these notations, applying the commutation relations, we obtain the following equalities:
$$
\langle z|z’\rangle = |\langle 0| W(-z)W(z’)|0\rangle| = 
|\langle 0|W(z’-z)|0\rangle|,
$$
and the first line in equality (\ref{1}) follows from the invariance of the vacuum vector.

So far we have used the commutativity and compactness of the group $K$. Now we use the maximality property.

Let $z\notin K$. Then there is such  $u\in K$ that  $\omega(z,u) \neq 1$. The following equalities are valid

\begin{multline}
\label{2}
\langle 0|W(z)|0\rangle = 	\langle 0|W(-u)W(u)W(z)|0\rangle = \\ =\omega(z,u) \langle 0|W(-u)W(z)W(u)|0\rangle = \omega(z,u) \langle 0|W(z)|0\rangle.
\end{multline}

In the first equality in the chain, we used the unitarity of the operators $W(u)$, in the second — the commutation relations, in the third — the invariance of the vacuum vector. It follows from the equalities (\ref{2}) that $\langle 0|W(z)|0\rangle = 0, \,z\notin K$. 

The uniqueness of the vacuum vector follows from the irreducibility of the representation. Indeed, let the invariant subspace not be one-dimensional, then we will choose in it a pair of orthogonal (invariant) vectors $|0_1\rangle$ and $|0_2\rangle$. A subspace in $\mathcal H_1$ spanned by vectors $W(z)|0_1\rangle, \,z\in F$ is a proper invariant subspace in $\mathcal H$, which contradicts the irreducibility of the representation. Lemma 1 is proved.

Lemma 1 has the following simple consequences.
\begin{corollary}
The Husimi function is constant in every coset in $F/K$ and is thus correctly defined on $F/K$.
\end{corollary}
\begin{corollary} For the Wehrl entropy , the formula is valid
$$
S^W_G(\rho) = -vol(K)\sum_{\alpha\in F/K}Q_\rho(\alpha)\log Q_\rho(\alpha).
$$
Here $vol(K)=\int_Kdz$.
\end{corollary}
\begin{corollary}
The lower bound of the Wehrl entropy is $0$, and it is achieved only on coherent states.
\end{corollary}
\begin{remark}
As noted above, the minimum of the Wehrl entropy for a group of real numbers is achieved on pure coherent states, that is, on Gaussian states. In the case of a group $G$ containing an open compact subgroup $H$, the coherent states are also Gaussian in the following sense. Let's consider the natural representation of CCR in $L^2(G)$. In this representation, the vacuum vector is the indicator function $\mathbb I_H$ of the subgroup $H$, that is, the Haar distribution of the compact subgroup. Such a distribution is Gaussian in the Bernstein sense (provided that $H$ is a Corwin group, that is, $2H=H$, \cite{Feldman}).
\end{remark}

To complete the proof of Theorem, we use the monotonicity property of the Wehrl entropy. Let the CCR representation for the $\mathbb R^n$ be realized in the space $\mathcal H_{\mathbb R^n}$, the representation for the $G_0$ is in the space $\mathcal H_{G_0}$. Then the CCR representation over $G=\mathbb R^n\times G_0$ is realized as the tensor product of the corresponding representations in the space $\mathcal H_{\mathbb R^n}\otimes \mathcal H_{G_0}$, and the inequalities are valid
$$
S^W_{\mathbb R^n\times G_0}(\rho)\geq S^W_{\mathbb R^n}(\rho_{\mathbb R^n})\geq n,
$$
where $\rho$ is the density operator in $\mathcal H_{\mathbb R^n}\otimes \mathcal H_{G_0}$, and $\rho_{\mathbb R^n}$  is the corresponding partial trace of the operator $\rho$. The theorem is proved.

\end{document}